# Unveiling New Mechanical Couplings in 3D Lattices: Axial-Bending and the Role of Symmetry Breaking

Dijia Zhong, Duo Qi, Jaehyung Ju

UM-SJTU Joint Institute, Shanghai Jiao Tong University

800 Dongchuan Road, Shanghai 200240, China

**Abstract**:

Mechanical couplings with symmetry breaking open up novel applications such as robotic metamaterials and directional mechanical signal guidance. However, most studies on 3D mechanical couplings have been limited to ad-hoc axial–twist designs due to a lack of comprehensive understanding of 3D non-centrosymmetry and chirality. Few theoretical methods exist to identify and quantify mechanical couplings in non-centrosymmetric and chiral lattices, typically relying on crystal physics (point group symmetry) and generalized constitutive equations. By extending symmetry breaking to mirror and inversion symmetries, we identify a broader range of mechanical couplings beyond axial-twist, such as axial-bending couplings. We develop a generalized 3D micropolar model of curved cubic lattices, encompassing both non-centrosymmetric achiral and chiral geometries, to quantify anisotropic physical properties and mechanical couplings as functions of curvature and handedness. Integrating point group symmetry operations with micropolar homogenized constitutive equations for curved cubic lattices, including mirror and inversion symmetry breaking, provides a clear design framework for identifying and quantifying anisotropic physical properties and mechanical couplings beyond axial-twist. This study uncovers a novel axial-bending coupling in non-centrosymmetric structures and highlights the weak correlation between chirality and both axial-bending and axial-twisting couplings. It also offers design guidelines for achieving multimodal couplings. The relationship between metamaterials' geometry and physical properties aligns with Neumann's principle. This work presents a robust framework for understanding mechanical couplings related to symmetry breaking and spatial anisotropy in metamaterial design, drawing an analogy to crystal physics and crystal chemistry.

**Keywords:** Symmetry Breaking, Mechanical Coupling, Axial-Bending, Chiral Lattices, Non-Centrosymmetric Structures, Point Group Symmetry, Micropolar Model

## 1. Introduction

Beyond the traditional focus of metamaterials on counterintuitive designs such as negative Poisson's ratio[1,2], negative thermal expansion[3,4], negative Young's modulus[5,6], negative bulk modulus[7,8], and non-reciprocal stiffness[9,10], recent interest has shifted towards mechanical couplings due to their potential applications in sensing, actuation of robotic metamaterials, and mechanical signal guidance in both time and space[11,12].

Symmetry theory from classical crystallography has long been used to relate atomic arrangements to physical properties. This framework can also help elucidate anisotropic properties and mechanical couplings in metamaterials. Various mechanical couplings have been studied in 2D structures by breaking lattice symmetry[11,13-15]. Chirality, in particular, has proven effective in generating axial-shear[16] (AS) and axial-rotation[17] (AR) couplings in 2D structures by breaking mirror symmetry while preserving N-fold rotational symmetry, thus inducing handedness[17,18]. Retaining rotational symmetry transforms axial deformation into symmetric (shear) and asymmetric (rotation) nodal responses. Breaking centrosymmetry in 2D structures can also facilitate axial-bending (AB) coupling[19], where the absence of inversion symmetry allows axial forces to induce bending moments[20].

In 3D structures, the complexity of design and mechanics increases due to additional degrees of freedom and more intricate symmetry considerations. Axial-twisting (AT) coupling has been a primary focus in the study of mechanical couplings in 3D lattices for decades[14,21-30]. Notably, 3D lattices exhibit four symmetry operations—translation, rotation, mirror, and inversion—resulting in 32 distinct point groups. The earliest AT coupling structure[14] was based on a four-fold rotational symmetry along three axes (Point group 432). Subsequent designs followed this template, replacing the windmill-like structure with spheres[21] or rigid cubic shapes[22,23]. Some designs[24,25] integrated spatially rotationally symmetric chiral elements extruded from each plane. Additionally, 3D structures using multiple materials have been designed within the same point group (432), introducing symmetry breaking to achieve AT coupling[30]. More recently, AT coupling designs have extended to other point groups, such as 422[26-28] and 4[29], which belong to non-centrosymmetric chiral structures.

However, few mechanical coupling effects beyond AT coupling have been explored in 3D lattices, primarily due to the lack of theoretical methods for identifying and quantifying mechanical couplings in both chiral and non-centrosymmetric lattices, supported by crystal physics (point group symmetry) and generalized constitutive equations. Analytical methods for quantifying mechanical couplings in given 3D lattice geometries remain limited. Therefore, in this study, we extend symmetry breaking to include both mirror and inversion operations in 3D structures to identify a broader range of mechanical couplings beyond axial-twist, such as axial-bending (AB) coupling. We develop a generalized 3D micropolar model of curved cubic lattices, incorporating both non-centrosymmetric achiral and chiral geometries, and quantify anisotropic physical properties and mechanical couplings as functions of curvature and handedness.

## 2. Chirality and Non-Centrosymmetry in 3D Metamaterials

Previous studies have often focused on localized chirality within a specific portion of a unit cell to analyze mechanical couplings, overlooking the overall symmetry of the structures. For instance, the pioneering design of an axial-twisting (AT) coupling structure[14] was based on a four-fold rotational symmetry along three axes, corresponding to point group 432. This configuration was initially thought to be chiral, which influenced subsequent designs. However, as shown in Figure 1a, the complete 3D unit cell is actually achiral and non-centrosymmetric.

In a 2D plane, the lattice face of a cubic cell exhibits chiral characteristics with R4 rotational symmetry, as it cannot be superimposed onto its mirror image under any transformation except reflection. Despite this, the 2D structure remains centrosymmetric due to the presence of R2 rotational symmetry. In 3D space, however, the entire 432 lattice can be mapped onto its mirror image through an inversion operation, making it achiral rather than chiral. This is because inversion inherently introduces a mirror component, violating the strict definition of chirality. Although the structure aligns with its mirror image under inversion, it does not map onto the original configuration, thus lacking true centrosymmetry.

Other studies also focused on local chirality in the 2D face of a unit cell when designing AT coupling structures[31]. This narrow perspective has led to misconceptions: Ha et al. [22] and Li et al. [23], for example, claimed that AT coupling can only occur in chiral metamaterials, which is incorrect. As detailed in Table S1 of the Supplementary Information (SI), there is no direct correlation between chirality and AT coupling.

Previous works on 3D cubic symmetry led to erroneous conclusions, such as: (i) AT coupling requires chirality in cubic symmetry[21,22,32], and (ii) AT coupling necessitates non-centrosymmetry in cubic symmetry[21,24]. However, cubic lattices, by definition[33], cannot exhibit chirality, as demonstrated by the achiral geometries in Figure 1b. Moreover, cubic symmetry is not inherently tied to non-centrosymmetry; it can encompass both centrosymmetric (e.g., point groups $m\bar{3}m$, $m\bar{3}$) and non-centrosymmetric (e.g., point groups 23, 432 and $\bar{4}3m$) configurations.

When researchers employed cubic symmetry to design geometries for AT coupling, their primary goal was often to simplify the elasticity tensor[32] while preserving non-centrosymmetry[24,34]. However, this approach oversimplifies the relationship between structural design and coupling effects. Non-centrosymmetric configurations lead to mechanical couplings due to the non-zero terms in the axial-rotation coupled elasticity tensor, often referred to as the **B** tensor[35]. Notably, a non-zero **B** tensor arising from non-centrosymmetry can produce a range of mechanical couplings beyond AT coupling[35], including Axial-Bending (AB), Shear-Twisting (ST), Shear-Bending (SB), Rotation-Twisting (RT), and Rotation-Bending (RR).

Distinguishing between chirality and non-centrosymmetry in 3D space is critical for avoiding ad-hoc design approaches and for advancing theory-driven designs in the metamaterials field. Figure 1b categorizes the 32 crystallographic point groups in 3D space, separating regions based on chirality and non-centrosymmetry. The geometric framework in Figure 1b uses the Wigner-Seitz cell[36] of the Bravais lattice, accounting for both point group symmetry (rotation, reflection, inversion, and improper rotation) and translational symmetry. The lattice geometries incorporate three shapes: cubes, regular octahedrons, and regular hexagonal prisms, as cubic lattices cannot generate point groups with three-fold or six-fold rotational symmetry. We design 3D geometries within this polyhedral framework by curving ligaments to create structures corresponding to specific point groups.

Centrosymmetry and chirality are mutually exclusive properties in 3D space. A centrosymmetric 3D lattice has an inversion center—a point through which every part of the structure has an equidistant, oppositely positioned counterpart. Since inversion in 3D can be decomposed into an R2 operation with mirror symmetry, structures exhibiting both centrosymmetry and chirality cannot exist in 3D crystallography, as illustrated in Figure 1b. Thus, the 32-point groups can be classified into three categories: (i) Non-centrosymmetric Chiral – structures that, after inversion, do not align with either their original or mirror image; (ii) Non-centrosymmetric Achiral – structures that align with their mirror image but not with the original configuration after inversion; and (iii) Centrosymmetric Achiral – structures that align with both their original and mirror image after inversion. Based on our previous work[35], potential mechanical couplings for these three domains are summarized in Table S1 of the Supplementary Information.

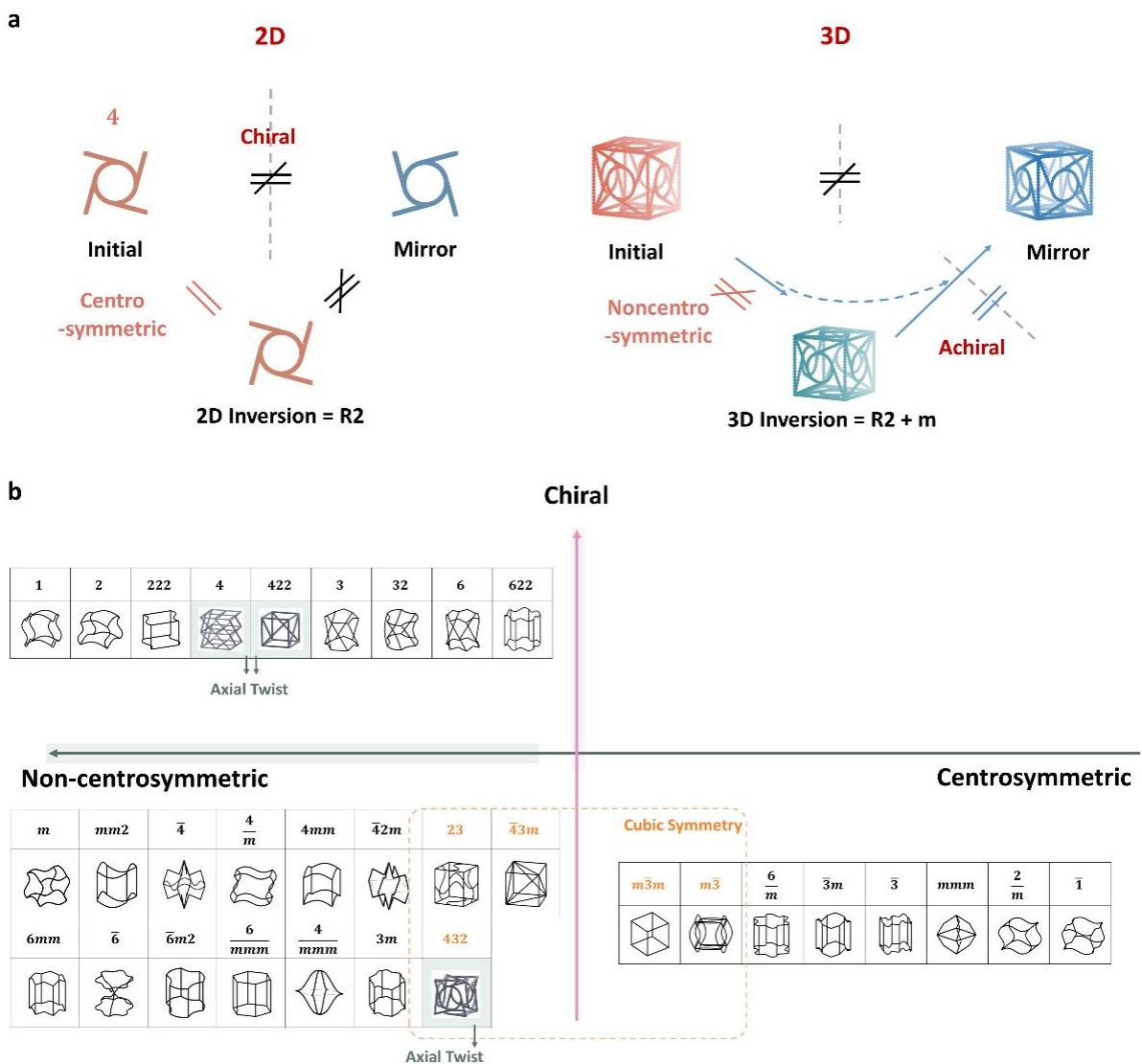

*Figure 1*. (a) *Illustration of chirality differences in 2D and 3D spaces, demonstrated using* the 432[14] *lattice and one of its surface geometries*; (b) *Classification of 3D symmetry point groups in the (non-) centrosymmetric and (a-)chiral domains with representative 3D lattices, including previous designs in point group* 4[15,29], 422[27] *and* 432[14].

## 3. Construction of non-centrosymmetric chiral and achiral structures

This work investigates 3D lattice structures with axial-bending (AB) coupling, extending beyond the traditional axial-torsion (AT) coupling. We theoretically demonstrate the AB coupling effect using constitutive equations and relate the lattice geometries to crystallography. Based on Table S1, which is constructed using a micropolar constitutive model and crystallographic point groups[35], lattices with point groups mm2 and 1 exhibit potential for AB coupling, with their geometries shown in Figure 2. The blue unit represents the smallest repeatable unit, while the red unit is designated as the Representative Volume Element (RVE). Selecting this RVE is crucial as it minimizes the overlap of components and ensures continuity across periodic joints[37-39], enhancing model clarity and reducing complexity[40].

Figure 2a illustrates a non-centrosymmetric achiral lattice belonging to the mm2 point group. This lattice features curved ligaments characterized by an undulation angle, $\beta$, forming a C shape in the $X_3$-direction and breaking rotational symmetry along the $X_1$ axis. Each ligament has a local coordinate system for its cross-section, denoted by lowercase labels, with dimensions $h \times t$ (height by thickness). When projected onto the $X_2X_3$ plane, the 2D geometry belongs to the m2 point group, aligning with prior research on 2D axial-bending coupling. The 3D lattice with mm2 symmetry introduces additional mirror planes: m$\perp X_1$ and m$\perp X_3$.

According to Table S1, the mm2 geometry potentially exhibits nine distinct mechanical couplings: Axial-Axial (AA), Axial-Bending (AB), Shear-Rotation (SR), Shear-Bending (SB), Shear-Twisting (ST), Rotation-Twisting (RT), Rotation-Bending (RB), Twisting-Twisting (TT), and Bending-Bending (BB). The characteristic C-shaped beams can be oriented in various configurations while maintaining the mm2 symmetry, as shown in Figure 2a. Each configuration supports axial-bending coupling along the curved segments.

Figure 2b presents a non-centrosymmetric chiral structure composed of three pairs of curved C beams. This design breaks global symmetry by incorporating additional curved beam pairs, placing the lattice within point group 1. The lack of further symmetry elements allows for all possible coupling effects. Different ligament orientations lead to various configurations, as depicted in Figure 2b, where four distinct examples are shown. Each configuration belongs to point group 1 but features differing orientations of inner segments. This variety in internal structure highlights the potential for customizing mechanical properties and coupling effects in chiral lattices.

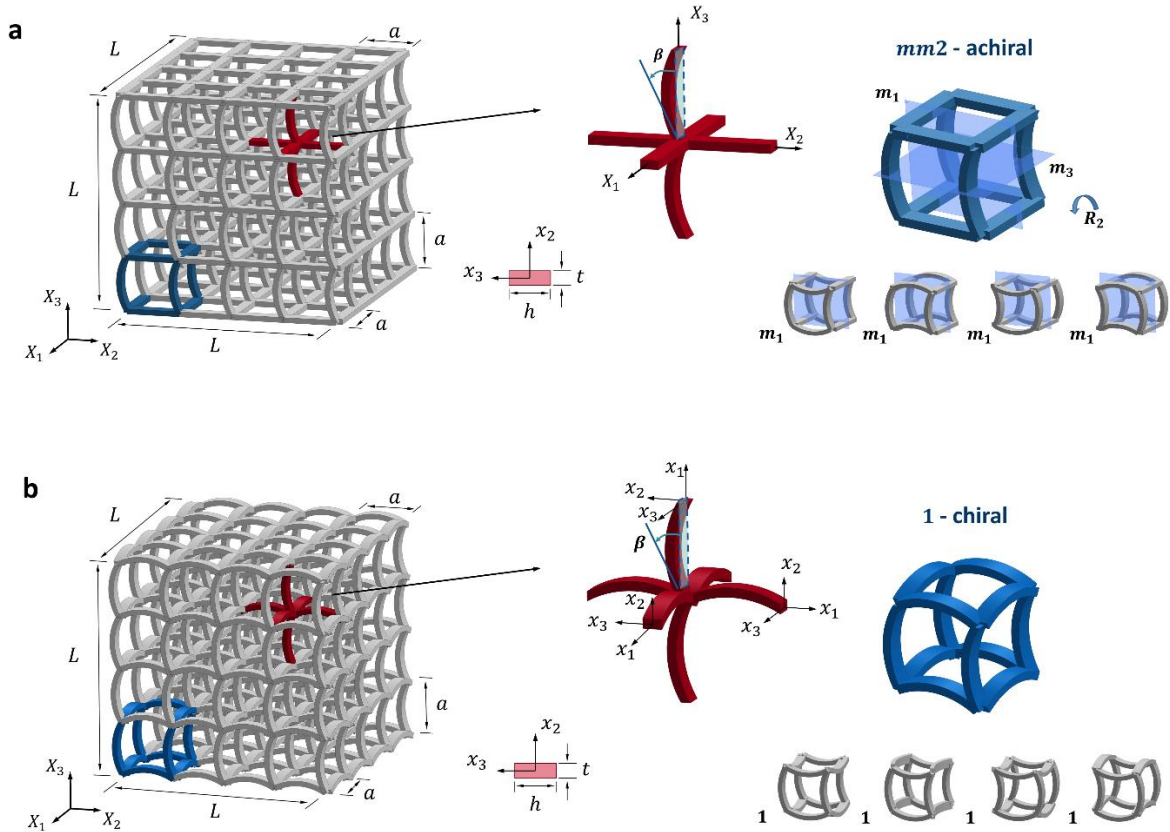

*Figure 2*. Construction of Non-Centrosymmetric Lattices: (a) Orthogonal tessellation of an achiral lattice belonging to the 'mm2' point group, shown as an $N \times N \times N$ configuration (L = Na, N = 4); (b) Orthogonal tessellation of a chiral lattice from the '1' point group, also shown as an $N \times N \times N$ configuration (with L = Na, N = 4).

## 4. Micropolar homogenization of 3D lattice structures

We employ micropolar elasticity to model the various mechanical loadings and couplings in 3D lattice structures[35]. Unlike Cauchy's elasticity, micropolar elasticity introduces three additional rotational degrees of freedom (DOFs) in the displacement field of a given particle[34]. Consequently, the displacement field of a particle is represented as $\mathbf{u} = \{u_1, u_2, u_3, \phi_1, \phi_2, \phi_3\}^T$, where $\phi_1, \phi_2$, and $\phi_3$ are the rotation angles with respect to the $X_1$, $X_2$, and $X_3$ axes, respectively.

The strain tensor $\varepsilon_{ij}$ and curvature tensor $\kappa_{ij}$, which are a function of the gradients of displacement $u_i$ and rotational angle $\phi_k$, are given by:

$$\varepsilon_{ij} = \frac{\partial u_i}{\partial x_j} - \epsilon_{ijk}\phi_k, \quad \kappa_{ij} = \frac{\partial \phi_i}{\partial x_j}, \qquad i, j, k = 1,2,3, \tag{1}$$

where $\epsilon_{ijk}$ is the permutation symbol and $\phi_k$ is the microrotation composed of macro-rotation, and nodal rotation[20]. Notably, $\varepsilon_{ij}$ is no longer symmetric due to the existence of the microrotation angle $\phi_k$.

The strain tensor has 18 independent components, which can be expressed in a vector form:

$$\varepsilon_{ij} = \{\varepsilon_{11}, \varepsilon_{22}, \varepsilon_{33}, \varepsilon_{23}, \varepsilon_{13}, \varepsilon_{12}, \varepsilon_{32}, \varepsilon_{31}, \varepsilon_{21}, \kappa_{11}, \kappa_{22}, \kappa_{33}, \kappa_{23}, \kappa_{13}, \kappa_{12}, \kappa_{32}, \kappa_{31}, \kappa_{21}\}^T$$

$$= \{\frac{\partial u_1}{\partial x_1}, \frac{\partial u_2}{\partial x_2}, \frac{\partial u_3}{\partial x_3}, \frac{\partial u_3}{\partial y} - \phi_1, \frac{\partial u_3}{\partial x_1} + \phi_2, \frac{\partial u_2}{\partial x_1} - \phi_3, \frac{\partial u_2}{\partial x_3} + \phi_1, \frac{\partial u_1}{\partial x_3} - \phi_2, \frac{\partial u_1}{\partial x_2} + \phi_3, \dots$$

$$\dots \frac{\partial \phi_1}{\partial x_1}, \frac{\partial \phi_2}{\partial x_2}, \frac{\partial \phi_3}{\partial x_3}, \frac{\partial \phi_2}{\partial x_3}, \frac{\partial \phi_1}{\partial x_3}, \frac{\partial \phi_1}{\partial x_2}, \frac{\partial \phi_3}{\partial x_2}, \frac{\partial \phi_3}{\partial x_1}, \frac{\partial \phi_2}{\partial x_1}\}^T \tag{2}$$

Notably, unlike the symmetric shear strain components in the classical Cauchy elasticity, the micropolar elasticity decouples the non-symmetric strain with the shear and local rotation[20].

Considering the constitutive relation $\sigma_{ij} = Q_{ijkl}\varepsilon_{kl}$, where $Q_{ijkl}$ is the fourth-order elasticity tensor, the stress tensor $\sigma_{ij}$ can be expressed with a vector form:

$$\sigma_{ij} = \{\sigma_{11}, \sigma_{22}, \sigma_{33}, \sigma_{23}, \sigma_{13}, \sigma_{12}, \sigma_{32}, \sigma_{31}, \sigma_{21}, m_{11}, m_{22}, m_{33}, m_{23}, m_{13}, m_{12}, m_{32}, m_{31}, m_{21}\}^T \tag{3}$$

The strain energy density $w$ of a lattice in the continuum model can be expressed as:

$$w = \frac{1}{2}\sigma_{ij}\varepsilon_{ij} = \frac{1}{2}\varepsilon_{ij}^T Q_{ijkl}\varepsilon_{kl} \tag{4}$$

To investigate the mechanical behavior of the 3D lattice structure, we use a homogenization scheme of a periodically tessellated RVE[41]. The elasticity tensor obtained through the micropolar homogenization of a 3D lattice structure can characterize the mechanical couplings.

The homogenization method for periodic lattices is well-established through the equivalence of strain energy between the micro and macro scales, as ensured by the Hill-Mandel condition[42-45]. This energy equivalence between the discretized lattice structure and the homogenized model is expressed as:

$$\int_V (\sigma_{ij}\varepsilon_{ij})dV = \frac{V}{2}\bar{\sigma}_{ij}\bar{\varepsilon}_{ij}, \tag{5}$$

where $\sigma_{ij}$ and $\varepsilon_{ij}$ are the actual stress and strain fields in the heterogeneous structures while $\bar{\sigma}_{ij}$ and $\bar{\varepsilon}_{ij}$ represent the homogenized average stress and strain, respectively. According to Hill's Lemma[46], this condition is satisfied when the boundary displacement is given by $u^i_{|\partial \Omega} = (\partial u^i/\partial x^j)x^j$ at every boundary node. Additionally, the displacement field $\mathbf{u^i}$ on the lattice joints is approximated using a Taylor expansion about the middle node.

$$\mathbf{u^i} = \mathbf{u^0} + \frac{\partial \mathbf{u^i}}{\partial \mathbf{x}} \Delta \mathbf{x} + \mathbf{o}\left((\Delta \mathbf{x^i})^2\right), \tag{6}$$

where $\mathbf{u^0}$ is the displacement of the middle node. Substituting Eq.(1) into Eq.(6), one can obtain the displacement and strain fields:

$$\begin{Bmatrix} u_1^i \\ u_1^i \\ u_1^i \\ \phi_1^i \\ \phi_2^i \\ \phi_3^i \end{Bmatrix} = \begin{Bmatrix} u_1^0 \\ u_1^0 \\ u_1^0 \\ \phi_1^0 \\ \phi_2^0 \\ \phi_3^0 \end{Bmatrix} + \begin{bmatrix} \bar{\varepsilon}_{11} & \bar{\varepsilon}_{21} - \bar{\phi}_3 & \bar{\varepsilon}_{31} + \bar{\phi}_2 \\ \bar{\varepsilon}_{12} + \bar{\phi}_3 & \bar{\varepsilon}_{22} & \bar{\varepsilon}_{32} - \bar{\phi}_1 \\ \bar{\varepsilon}_{13} - \bar{\phi}_2 & \bar{\varepsilon}_{23} + \bar{\phi}_1 & \bar{\varepsilon}_{33} \\ \bar{\kappa}_{11} & \bar{\kappa}_{12} & \bar{\kappa}_{13} \\ \bar{\kappa}_{21} & \bar{\kappa}_{22} & \bar{\kappa}_{11} \\ \bar{\kappa}_{31} & \bar{\kappa}_{32} & \bar{\kappa}_{33} \end{bmatrix} \begin{Bmatrix} x_1^i \\ x_2^i \\ x_3^i \end{Bmatrix} + \mathbf{o}\left((\Delta x^i)^2\right), \tag{7}$$

where $\bar{\varepsilon}_{ij}$, $\bar{\phi}_i$, and $\bar{\kappa}_{ij}$ are the strain tensor, rotational angle, and curvature tensor of the middle node. The strain energy $W$ of RVE of a 3D lattice structure can be written as

$$W = \frac{V}{2} \bar{\sigma}_{ij} \bar{\varepsilon}_{ij} = \frac{1}{2} \mathbf{d}^T \cdot \mathbf{K} \cdot \mathbf{d}, \tag{8}$$

where the displacement vector $\mathbf{d_i} = \{\mathbf{u_1^i}, \mathbf{u_2^i}\}^T$ has a dimension of $12 \times 1$, aligning with the displacement field of the homogenized continuum. The stiffness tensor $\mathbf{K}$ for an RVE is has $12 \times 12$ components where three rotational moments of inertia correspond to two bending modes and one twisting mode, which confirms the suitability of this beam model for accurately capturing different deformation patterns. One can obtain the strain energy $W$ in Equation (8) using the finite element-based discretization method (see the detail in **Section S2** of SI).

Combing Equations (4) with (6), one can obtain elasticity tensor $Q_{ijkl}$ by

$$Q_{ijkl} = \frac{\partial^2 w}{\partial \bar{\varepsilon}_{ij} \partial \bar{\varepsilon}_{kl}} \tag{9}$$

From materials and geometric parameters of $E = 9.3 MPa$, $v = 0.3$, $t/L = 0.1$, $\theta = 30°$, we can obtain the elasticity tensors $\mathbf{Q}_{mm2}$ and $\mathbf{Q}_1$ for non-centrosymmetric achiral and non-centrosymmetric chiral structures, respectively. Notably, the $B_{35}$ component of $\mathbf{Q}_{mm2}$ and the $B_{35}$, $B_{26}$, and $B_{16}$ components of $\mathbf{Q}_1$ are non-zero in Equation (10). All numerical values of the components of the two elasticity tensors $\mathbf{Q}_{mm2}$ and $\mathbf{Q}_1$ are shown in SI.

$$\begin{Bmatrix}\sigma_{11}\\\sigma_{22}\\\sigma_{33}\\\sigma_{23}\\\sigma_{13}\\\sigma_{12}\\\sigma_{32}\\\sigma_{31}\\\sigma_{21}\\m_{11}\\m_{22}\\m_{33}\\m_{23}\\m_{13}\\m_{12}\\m_{32}\\m_{31}\\m_{21}\end{Bmatrix} = \begin{bmatrix} C_{11} & C_{12} & C_{13} & C_{14} & C_{15} & C_{16} & C_{17} & C_{18} & C_{19} & B_{11} & B_{12} & B_{13} & B_{14} & B_{15} & B_{16} & B_{17} & B_{18} & B_{19}\\ & C_{22} & C_{23} & C_{24} & C_{25} & C_{26} & C_{27} & C_{28} & C_{29} & B_{21} & B_{22} & B_{23} & B_{24} & B_{25} & B_{26} & B_{27} & B_{28} & B_{29}\\ & & C_{33} & C_{34} & C_{35} & C_{36} & C_{37} & C_{38} & C_{39} & B_{31} & B_{32} & B_{33} & B_{34} & B_{35} & B_{36} & B_{37} & B_{38} & B_{39}\\ & & & C_{44} & C_{45} & C_{46} & C_{47} & C_{48} & C_{49} & B_{41} & B_{42} & B_{43} & B_{44} & B_{45} & B_{46} & B_{47} & B_{48} & B_{49}\\ & & & & C_{55} & C_{56} & C_{57} & C_{58} & C_{59} & B_{51} & B_{52} & B_{53} & B_{54} & B_{55} & B_{56} & B_{57} & B_{58} & B_{59}\\ & & & & & C_{66} & C_{67} & C_{68} & C_{69} & B_{61} & B_{62} & B_{63} & B_{64} & B_{65} & B_{66} & B_{67} & B_{68} & B_{69}\\ & & & & & & C_{77} & C_{78} & C_{79} & B_{71} & B_{72} & B_{73} & B_{74} & B_{75} & B_{76} & B_{77} & B_{78} & B_{79}\\ & & & & & & & C_{88} & C_{89} & B_{81} & B_{82} & B_{83} & B_{84} & B_{85} & B_{86} & B_{87} & B_{88} & B_{89}\\ & & & & & & & & C_{99} & B_{91} & B_{92} & B_{93} & B_{94} & B_{95} & B_{96} & B_{97} & B_{98} & B_{99}\\ & & & & & & & & & D_{11} & D_{12} & D_{13} & D_{14} & D_{15} & D_{16} & D_{17} & D_{18} & D_{19}\\ & & & & & & & & & & D_{22} & D_{23} & D_{24} & D_{25} & D_{26} & D_{27} & D_{28} & D_{29}\\ & & & & \text{symmetric} & & & & & & & D_{33} & D_{34} & D_{35} & D_{36} & D_{37} & D_{38} & D_{39}\\ & & & & & & & & & & & & D_{44} & D_{45} & D_{46} & D_{47} & D_{48} & D_{49}\\ & & & & & & & & & & & & & D_{55} & D_{56} & D_{57} & D_{58} & D_{59}\\ & & & & & & & & & & & & & & D_{66} & D_{67} & D_{68} & D_{69}\\ & & & & & & & & & & & & & & & D_{77} & D_{78} & D_{79}\\ & & & & & & & & & & & & & & & & D_{88} & D_{89}\\ & & & & & & & & & & & & & & & & & D_{99}\end{bmatrix}\begin{Bmatrix}\varepsilon_{11}\\\varepsilon_{22}\\\varepsilon_{33}\\\varepsilon_{23}\\\varepsilon_{13}\\\varepsilon_{12}\\\varepsilon_{32}\\\varepsilon_{31}\\\varepsilon_{21}\\\kappa_{11}\\\kappa_{22}\\\kappa_{33}\\\kappa_{23}\\\kappa_{13}\\\kappa_{12}\\\kappa_{32}\\\kappa_{31}\\\kappa_{21}\end{Bmatrix}$$

where the elastic tensor producing an axial-bending coupling, $\mathbf{Q}_{ab} = \begin{bmatrix} B_{14} & B_{15} & B_{16} & B_{17} & B_{18} & B_{19}\\ B_{24} & B_{25} & B_{26} & B_{27} & B_{28} & B_{29}\\ B_{34} & B_{35} & B_{36} & B_{37} & B_{38} & B_{39}\end{bmatrix}$ (10)

The decoupled constitutive relation from the generalized micropolar theory for a lattice with the 'mm2' point group indicates nine potential mechanical couplings: AA, AB, SR, ST, SB, RT, RB, TT, and BB (as listed in Table S1 of the Supplementary Information). However, for a specific geometry—the non-centrosymmetric achiral lattice shown in Figure 2a—the non-zero components from Equation (10) reveal only two dominant mechanical couplings: AB and SB. The AT coupling terms are zero, and the remaining coupling terms are negligible in $\mathbf{Q}_{mm2}$. It is important to note that the potential couplings listed in Table S1 are derived from the decoupled micropolar constitutive relation based on Neumann's principle, which is a necessary but not sufficient condition for determining the actual mechanical couplings of a specific geometry.

For a lattice with the '1' point group, the number of potential mechanical couplings is 15, as indicated in Table S1. However, the non-centrosymmetric chiral lattice with the '1' point group shown in Figure 2b exhibits only 12 mechanical couplings according to $\mathbf{Q}_1$; three coupling terms—AS, AR, and AT—are absent.

## 5. Anisotropic mechanical properties

By selectively breaking specific symmetries within cubic lattices while preserving mirror symmetry, we introduce pronounced anisotropy in both the effective moduli and mechanical couplings. We evaluate the effective moduli of non-centrosymmetric achiral ('mm2') and chiral ('1') lattices, assuming a fixed slenderness ratio of $t/L = 0.1$ with $L = 10mm$. Figure 3 shows the directional mechanical properties of these lattices for various undulation angles: $\beta = 10°, 20°$, and $30°$. To visualize these properties, we transform the fourth-rank compliance tensor $\mathbf{S} = (\mathbf{Q}^{-1})$ into spherical coordinates[34]. The effective terms, computed using dyadic bases in the spherical coordinate system, are expressed as follows:

$$E^* = (\mathbf{e}_R\mathbf{e}_R : \mathbf{S} : \mathbf{e}_R\mathbf{e}_R)^{-1} \tag{10}$$

$$\nu^* = -E^*\mathbf{e}_R\mathbf{e}_R : \mathbf{S} : \mathbf{e}_\phi\mathbf{e}_\phi \tag{11}$$

$$G^* = \frac{1}{2}\left[\frac{\sqrt{2}}{2}(\mathbf{e}_R\mathbf{e}_\phi + \mathbf{e}_\phi\mathbf{e}_R) : \mathbf{S} : \frac{\sqrt{2}}{2}(\mathbf{e}_R\mathbf{e}_\phi + \mathbf{e}_\phi\mathbf{e}_R)\right]^{-1} \tag{12}$$

$$B^* = \frac{1}{3}(\mathbf{e}_R\mathbf{e}_R : \mathbf{S} : \mathbf{e}_R\mathbf{e}_R + \mathbf{e}_R\mathbf{e}_R : \mathbf{S} : \mathbf{e}_\phi\mathbf{e}_\phi + \mathbf{e}_R\mathbf{e}_R : \mathbf{S} : \mathbf{e}_\theta\mathbf{e}_\theta)^{-1} \tag{13}$$

$$\zeta^*_{ab} = E^*\mathbf{e}_R\mathbf{e}_R : \mathbf{S} : \mathbf{A} \tag{14}$$

where $E^*$, $G^*$, and $B^*$ denote the effective Young's modulus, shear modulus, and bulk modulus, respectively. The effective Poisson's ratio is represented by $\nu^*$, while $\zeta^*_{ab}$ indicates the axial-bending coupling ratio. The unit basis vectors $\mathbf{e}_R$, $\mathbf{e}_\theta$, and $\mathbf{e}_\phi$ correspond to the radial distance $R$, polar angle $\theta$, and azimuthal angle $\phi$ in spherical coordinates.

To fully capture the mechanical behavior of these lattices, we introduce a second-rank anti-identity tensor $\mathbf{A}$ when deriving $\zeta^*_{ab}$ in Equation (14). The tensor $\mathbf{A}$ has zero diagonal elements and off-diagonal elements equal to one:

$$\mathbf{A} = \begin{bmatrix} 0 & 1 & 1 \\ 1 & 0 & 1 \\ 1 & 1 & 0 \end{bmatrix}, \tag{15}$$

The anti-identity tensor $\mathbf{A}$ is employed because a uniaxial load applied to the lattice does not generate a curvature gradient in the plane perpendicular to the load direction (as elaborated in Section 5 of the Supplementary Information). Instead, the off-diagonal terms capture the coupling between the axial loading and the resulting curvature response.

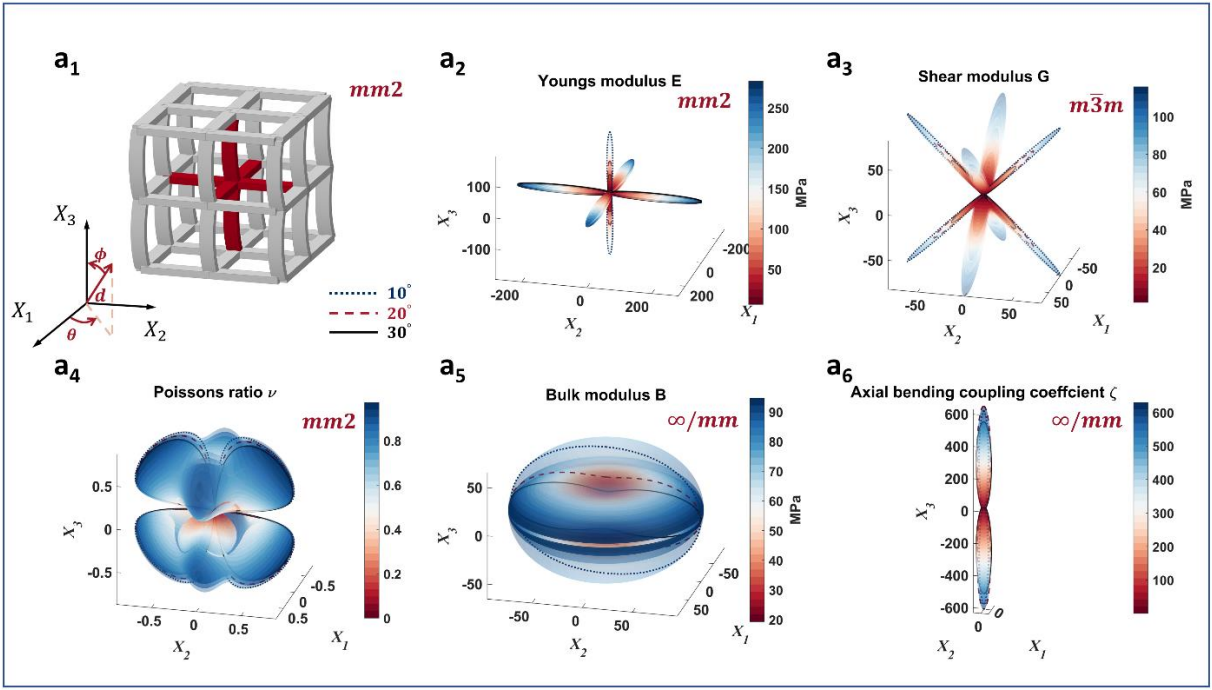
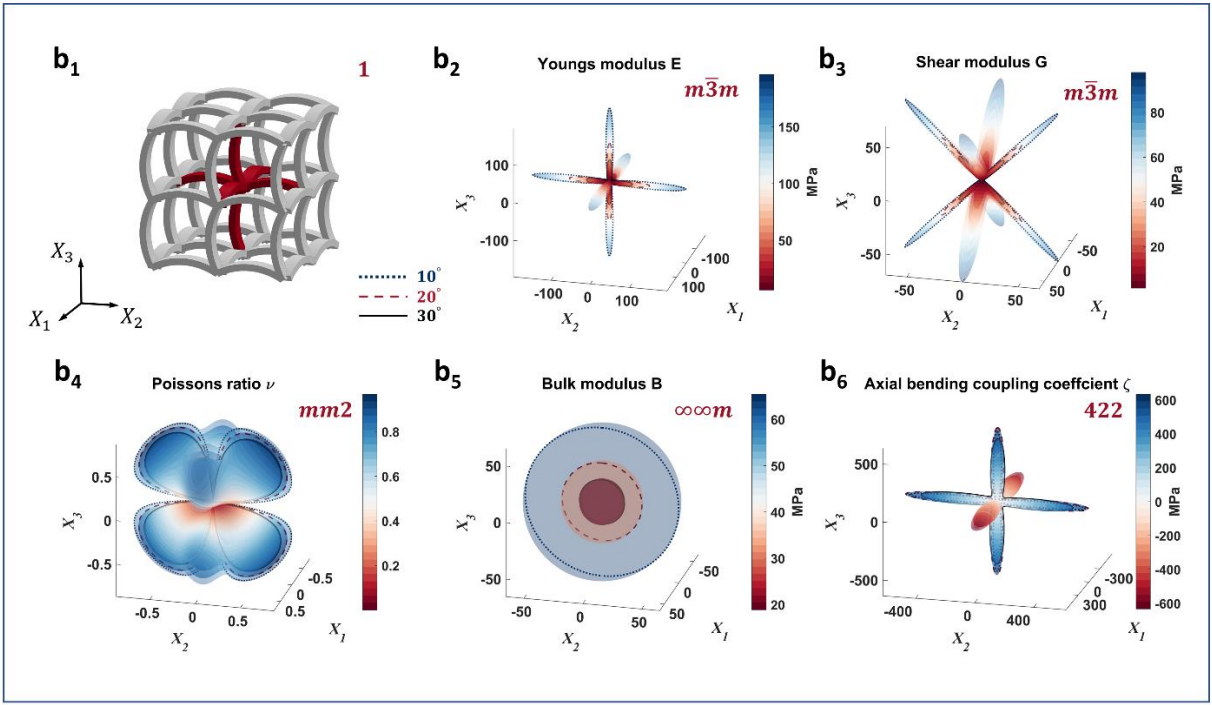

Figure 3. Anisotropic mechanical properties of (a) non-centrosymmetric achiral ('mm2') and (b) non-centrosymmetric chiral ('1') lattices, along with their point groups.

The pattern of anisotropic properties shown in Figure 3 is analyzed using Neumann's principle [47], which establishes the fundamental relationship between physical properties and structural symmetry. According to this principle, the symmetry point group of any physical property must be a subgroup of the geometric point group. Additionally, the Curie symmetry group is applied due to the continuous directional plot of properties in three-dimensional space. Relationships between subgroups and supergroups of various point groups are detailed in Section S2 of the SI. A pronounced anisotropy is observed in both non-centrosymmetric achiral ($mm2$) and chiral (1) lattices, with a focus on the symmetry point group of the directional properties.

Young's modulus shows significant anisotropy along three principal directions: both lattice types exhibit peaks at $\theta = k_1\pi/2$ and $\varphi = k_2\pi$, where $k_1 = 0,1,2,3$ and $k_2 = 0,1$. Increasing the curvature of the beam components reduces the modulus. The non-centrosymmetric achiral ($mm2$) lattice exhibits anisotropy with reduced modulus in the $X_3$ direction due to the curved beams in that direction. In contrast, the non-centrosymmetric chiral (1) lattice, shown in Figure 3b1, with curved beams in all three directions, yields uniform modulus magnitudes. A greater β value reduces the magnitude of Young's modulus. The modulus pattern of the non-centrosymmetric achiral lattice corresponds to the $mm2$ point group, as shown in Figure 3a2, while the chiral lattice's pattern follows the m3m point group, as seen in Figure 3b2.

Unlike Young's modulus, which can be independently tuned in specific directions, the shear modulus exhibits equal magnitudes along the protruding directions due to mutual interactions between neighboring beams. Both the non-centrosymmetric achiral ($mm2$) and chiral (1) lattices display the same shear modulus symmetry, represented by the $m\bar{3}m$ point group, as depicted in Figures 3a2 and 3b2. Maximum shear stresses occur on the diagonal planes where $\theta \approx k_1\pi/2$ and $\varphi \approx (2k_2 - 1)\pi/4$, with $k_1 = 0,1,2,3$, $k_2 = 0,1$. The achiral lattice ('$mm2$') produces approximately 18.3% higher shear modulus than the chiral lattice due to the presence of straight beams surrounding the curved beam, enhancing stiffness. Increased curvature (greater β) reduces the shear modulus.

The bulk modulus patterns differ between the two lattices. The chiral (1) lattice displays an isotropic property with an $\infty\infty m$ Curie group symmetry, as shown in Figure 3b5, while the achiral ($mm2$) lattice shows anisotropy in the $X_3$ direction due to its undulated structure, producing an $\infty/mm$ point group pattern, as shown in Figure 3a5. Higher beam curvature lowers the bulk modulus magnitude.

Both lattices exhibit the same symmetry pattern ($mm2$) for Poisson's ratio, with low directional sensitivity and a positive maximum value close to 1. However, significant variation is observed near principal loading directions, leading to a zero Poisson's ratio, as shown in Figures 3a4 and 3b4. Previous studies on 2D chiral and achiral square lattices showed similar Poisson's ratio patterns, except for a tilted pattern in chiral lattices due to axial-shear coupling and dual (positive and negative) Poisson's ratios in achiral lattices[19].

For the achiral lattice ($mm2$), the symmetry pattern for Young's modulus and Poisson's ratio matches the $mm2$ point group, as shown in Figure 3a. In contrast, the shear modulus belongs to the higher-ranked $m\bar{3}m$ point group ($m \perp Z_1, 3 \parallel [1,1,1], m \perp [1\ 1\ 0]$), indicating that it possesses higher symmetry compared to Young's modulus and Poisson's ratio. The bulk modulus and axial bending (AB) coupling coefficients align with the '$\infty/mm$' Curie group, exhibiting mirror symmetry about the $X_1$ and $X_2$ axes and infinite rotational symmetry around the $X_3$ axis, a feature absent in the $mm2$ point group.

The chiral lattice (1) has the lowest symmetry among the structures, as indicated in the hierarchical symmetry map in Figure S1 of the SI. Interestingly, this low-symmetry geometry can still produce highly symmetric physical properties, such as the bulk modulus ($\infty\infty m$) shown in Figure 3b5. The symmetry elements of all physical properties ($\infty\infty m$, $m\bar{3}m$, and 422) of the chiral lattice are higher than the point group symmetry (1), demonstrating that the relationship between geometric structure and physical properties in metamaterials adheres to Neumann's principle.

While traditional crystal physics does not address mechanical coupling, our analysis identifies the symmetry elements of unique mechanical coupling coefficients, such as axial bending, in metamaterials. Notably, the mechanical coupling effects in these structures also follow Neumann's principle.

## 6. Validation of axial-bending coupling with experiments and FE simulations

We implemented the nodal displacement analysis of the homogenized 3D lattice model (Figure 2a) using a custom MATLAB code (see Supporting Information). To validate the axial-bending coupling of the micropolar homogenized model, we performed finite element (FE) simulations using ABAQUS/Standard, modeling the lattices with Euler-Bernoulli beam elements (B31 in ABAQUS). The lattice geometry was set as $\beta = 30°$, $t/a = 0.1$, and $L = 90mm$, with varying cell sizes. For the base material, thermoplastic polyurethane (TPU), we assigned a modulus of $E = 9.3\ MPa$ and a Poisson's ratio of $\nu = 0.3$.

Figure 4 compares the nodal displacements from the analytical model, FE simulations, and experimental results, showing the joint positions after deformation in the FE simulations. We 3D printed lattice samples using TPU with a commercial extrusion printer (Bambu Lab X1) and measured the nodal displacements using the digital image correlation (DIC) method.

In Figure 4a, we observe good agreement between the homogenized model and the FE simulations under a small strain of 0.1%, adhering to the linear continuum assumption. However, some discrepancies arise due to geometric nonlinearity, manifesting as the well-known "size effect." The deformation pattern is influenced by both geometric incompatibility and the size effect of homogenization. As the number of lattice cells increases, the middle span of the bending curve tends to flatten, likely due to incompatible rotation angles caused by the tessellated curved beams. Figure 4c illustrates how vertical compression induces counteracting moments on adjacent curved beams, resulting in horizontal deformation.

The size effect becomes more pronounced when scaling up from the representative volume element (RVE) to larger assemblies. This effect arises from the misalignment between local and global rotation centers, a phenomenon discussed in detail as "screw theory" by Xu et al [48]. Some approaches aim to mitigate the size effect on coupling outputs by reducing the number of symmetry axes in the global assembly to minimize interface conflicts[15,27,48]. However, unit cells with mirror symmetry inherently generate additional symmetry planes during tessellation, preventing full alignment along a single mirror plane. In contrast, units with pure rotational symmetry (as in cylindrical tessellation) can be arranged to share a common rotation axis[15]. Therefore, axial-bending (AB) lattices with mirror symmetry will inevitably exhibit size effects as the number of lattice cells increases perpendicular to the symmetry plane.

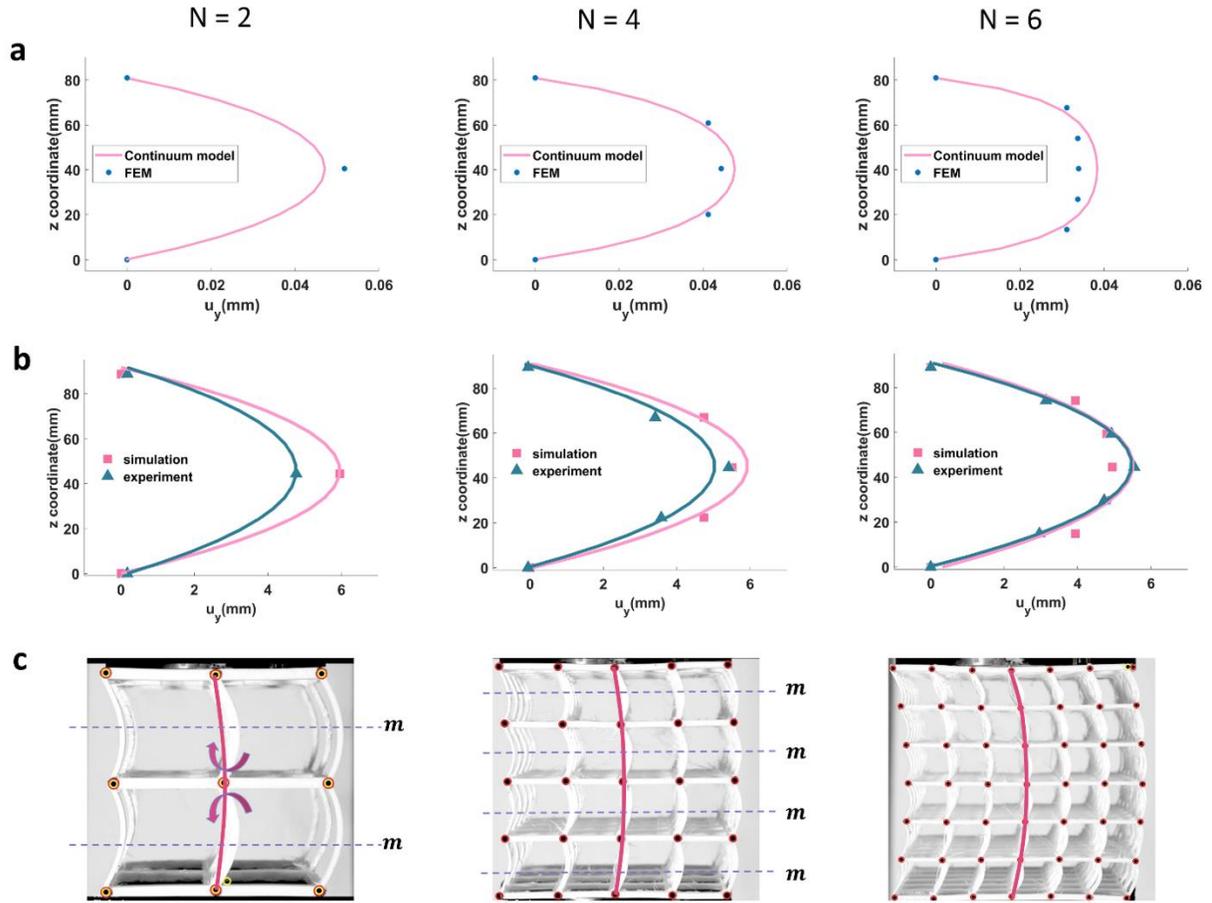

*Figure 4. Bending deformation along the centerline nodes $(X = L/2, Y = L/2)$ of the 'mm2' lattice for different lattice sizes of $N = 2, 4, 6$. **a** Comparison of FE simulations and the homogenized model under an axial strain of -0.1%; **b** Joint displacements from simulations and experimental measurements; **c.** Deformed configuration observed in the experiment for an axial strain of $-3\%$.*

Figure 4b shows the lattice deflection observed in experiments compared to FE simulations under a larger axial strain (-3%). Unlike the FE simulations for larger cell numbers ($N = 6$), the experimental results do not exhibit a flattening effect in the middle region. Instead, the bending curve is symmetric and smooth, indicating the absence of buckling and suggesting a predictable deformation response as described by the elastic tensor **Q**. The experimental bending deformation curve is shown in Figure 3c.

In contrast to the FE simulations, which display a size effect where the maximum bending deflection decreases from 5.96 mm to 4.94 mm as the number of cells $N$ increases, the experiment shows a slight increase in deflection from 4.77 mm to 5.54 mm. As the number of cells increases, the beam thickness decreases proportionally ($t/a = 0.1$, where $a = L/N$), leading to thinner beams and, consequently, a greater overall bending deformation.

## 7. Multimodal mechanical coupling effects

Multimodal mechanical coupling is crucial for selectively guiding directional load transfer and dissipating mechanical energy across multiple modes, including axial, bending, shear, and twist. To design such couplings, including axial-bending (AB) and others, we explore lattice geometries from point groups $1, 2, \boldsymbol{m}, \boldsymbol{mm2}, \boldsymbol{4}, \bar{4}, \boldsymbol{4mm}, 3, 32, 3m, 6, \bar{6}, 6mm$, and $\bar{6}m2$, as listed in Table S1, which show potential for AB coupling [35].

In Figure 5a, we present a non-centrosymmetric achiral lattice belonging to the $4mm$ point group. Compared to the $mm2$ lattice in Figure 2a, an additional pair of curved members introduces a mirror symmetry ($m \perp X_1$), resulting in double AB coupling for axial loads in the $X_1$ and $X_3$ directions. By replacing the straight beams with curved ones, one can create a lattice with triple AB coupling, breaking the overall mirror symmetry but maintaining independent AB couplings along the three orthogonal directions, as seen in the non-centrosymmetric chiral lattice of point group '1' in Figure 2b.

Figure 5b illustrates a non-centrosymmetric achiral lattice from the $mm2$ point group, exhibiting dominant axial-axial (AA) and AB couplings. Unlike the lattice in Figure 2a, this design produces a negative Poisson's ratio. The inclusion of inner-centered curved members on the $X_2X_3$ plane leads to a negative Poisson's ratio under loading in the $X_2$ and $X_3$ directions [35].

Recognizing that S-shaped curved beams embedded in tetragons induce axial shear (AS) coupling [16], this concept can be extended to a 3D lattice design. In Figure 5c, we introduce a non-centrosymmetric achiral lattice belonging to the 'm' point group, featuring four S-shaped curves on the front and rear $X_2X_3$ planes and four C-shaped curves on the $m \perp X_1$ plane. This configuration results in AS coupling along the $X_2X_3$ planes under axial loads in the $X_2$ and $X_3$ directions, while also providing AB coupling on the $X_1X_3$ plane for axial loads in the $X_1$ direction.

Unlike the independent multimodal couplings in Figures 5a-5c, it is possible to design coupled mechanical responses that are interdependent across multiple directions. For example, a 3D structure can be designed with dominant axial-torsion (AT) and AB couplings. By selecting the '4' point group, a non-centrosymmetric chiral lattice, we can construct an S-shaped beam on the (1,1,0) plane and rotate it by 90° along the $X_3$ axis (4 ∥ $X_3$). By connecting the remaining nodes with C-shaped beams in a cubic lattice arrangement, we obtain a structure with dependent AT and AB couplings. For loading along the $X_3$ axis, shear deformation on the (1,1,0) plane is coupled with bending deformation on the (1,0,0) plane.

It is important to note that the multimodal coupling effects in Figures 5a, 5b, and 5c are independent. For instance, the AB coupling in the $X_1$ direction does not affect the AB coupling in the $X_2$ direction in Figure 5a. In contrast, the multimodal coupling in Figure 5d is dependent, as axial loading in the $X_2$ direction for bending can induce additional coupling in the (1,1,0) plane. These independent and dependent couplings can be validated through the elasticity tensor **Q** in Equation (10). Independent multimodal couplings are represented by distinct rows in the **Q** tensor.

Additionally, due to the reciprocal nature[35] of the coupling effects, non-axial inputs can generate other types of couplings. For example, bending-axial (BA) and shear-axial (SA) couplings may occur in lattice structures that already exhibit AB and AS couplings.

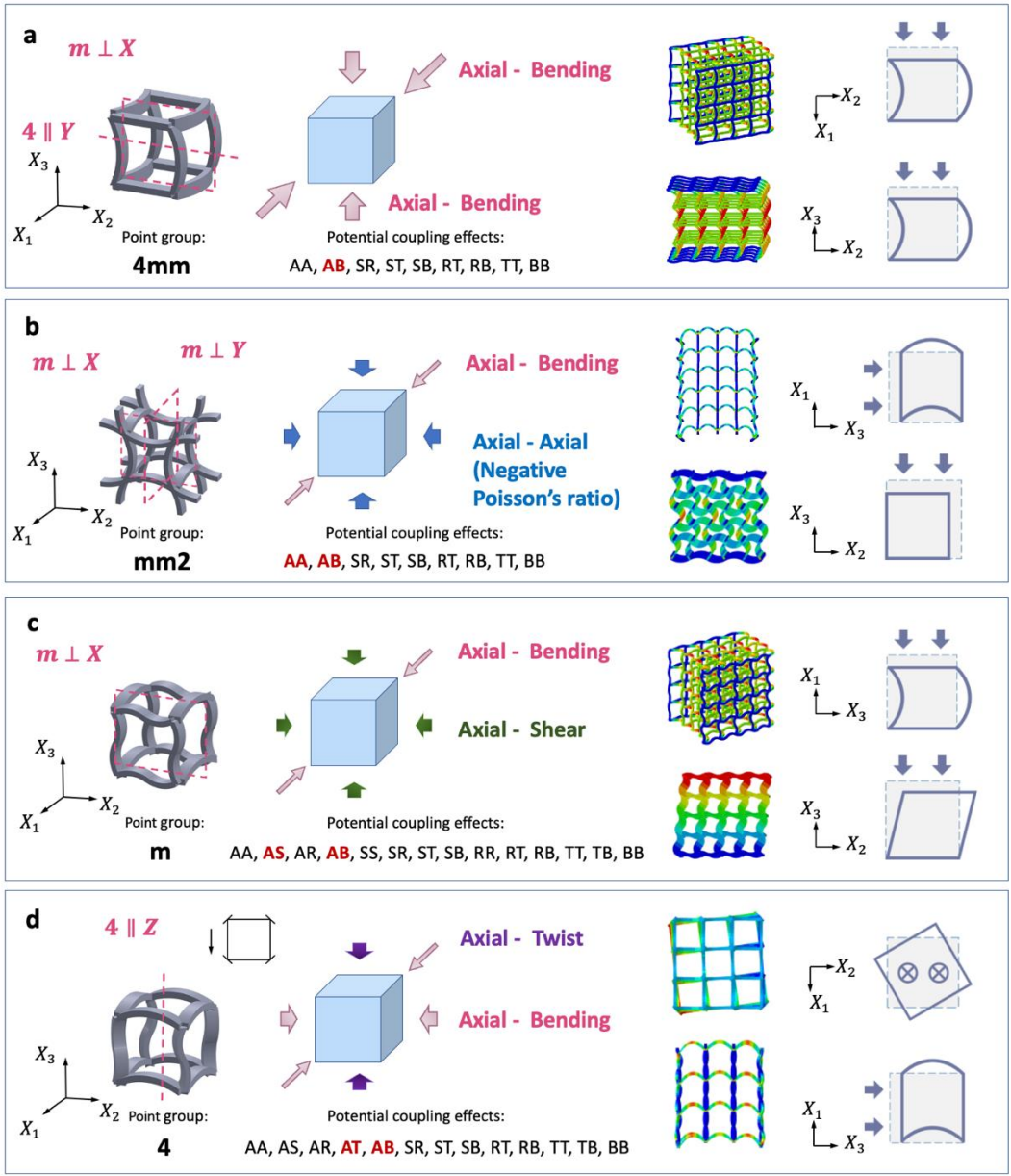

*Figure 5. Multimodal couplings in curved cubic lattice designs; **a.** a non-centrosymmetric achiral lattice belonging to the **4mm** point group, featuring one mirror symmetry - $m \perp X_1$ and one 4-fold rotation symmetry ($4 \parallel X_2$); **b.** a non-centrosymmetric achiral lattice belonging to the **mm2** point group with two mirror symmetries ($m \perp X_1$ and $m \perp X_2$); **c.** a non-centrosymmetric achiral lattice belonging to the **m** point group, exhibiting a single mirror symmetry ($m \perp X_1$); **d.** a non-centrosymmetric chiral lattice from the **4** point group, characterized by a four-fold rotational symmetry ($4 \parallel X_3$).*

## 8. Discussion

The design of 3D lattice metamaterials incorporating mechanical coupling effects not only facilitates the discovery of materials with significant negative Poisson's ratios [9,6] but also enhances our understanding of motion transformation across different modes [15,49]. However, due to limited exploration of geometric symmetries and weak connections to generalized constitutive models, existing research on non-centrosymmetric structures—often labeled as "chiral"—has primarily focused on axial twisting couplings in 3D [14,21,23,30]. To address the limitations of current ad-hoc design approaches, we propose a generalized framework that captures a broader range of mechanical couplings in 3D lattice structures.

Mirror and inversion symmetry operations, though developed as independent concepts, exhibit overlapping features when mapping the geometries of 32-point group elements. Thus, classifying these 32-point group elements on a 2D map (Figure 1) with chiral and non-centrosymmetric characteristics effectively identifies the impact of symmetry breaking on physical properties. Notably, the metamaterial community often conflates symmetry-breaking effects with chirality alone [19,2], overlooking the broader geometric symmetry context in 3D space.

Current 3D micropolar homogenization methods are typically limited to straight beam elements [50] or become computationally intensive when extended to full finite element models [51]. Simplified elasticity tensors often assume cubic [21] or orthogonal [32], symmetry, which can lead to inaccurate representations of lattice symmetries and fail to capture the anisotropic and coupling responses of metamaterials [52].

We introduce a generalized 3D micropolar homogenization method incorporating a complete elasticity tensor without restrictive symmetry assumptions. We designed non-centrosymmetric achiral ('mm2' point group) and chiral ('1' point group) lattices to examine anisotropic behavior and mechanical couplings beyond axial twisting. Our generalized model accurately predicts the axial-bending coupling of architected materials, showing good agreement with FE simulations and experiments. The analytical model also identifies the optimal geometric parameters for maximizing the coupling coefficient, such as an undulation angle of $\sim 10°$ for $t/L = 0.1$, while quantifying the anisotropic moduli and Poisson's ratio, which are correlated with symmetric point groups.

This work extends the application of Neumann's principle beyond traditional crystallography. By integrating both point group and Curie group principles, we establish a consistent hierarchical relationship between geometric symmetry and property symmetry. This correlation enables inverse design strategies for achieving target anisotropic properties and mechanical couplings. The multimodal couplings illustrated in Figure 5 have potential applications in developing robotic metamaterials, guided signal transfer in mechanical computing devices, and the physical intelligence of soft robots.

## 9. Conclusion

By integrating point group symmetry operations with micropolar homogenized constitutive equations for curved cubic lattices, including considerations of mirror and inversion symmetry breaking, we establish a comprehensive design framework to identify and quantify anisotropic physical properties and mechanical couplings beyond axial twisting. This study reveals a novel axial-bending coupling in non-centrosymmetric structures and highlights a weak correlation between chirality and both axial-bending and axial-twisting couplings, providing clear design guidelines for achieving multimodal couplings.

The analytical model determines the optimal geometric parameters for maximizing the coupling coefficient and effectively quantifies anisotropic moduli and Poisson's behaviors, which are associated with symmetric point groups. Our findings confirm that the relationship between the geometry of metamaterials and their physical properties, including mechanical couplings, adheres to Neumann's principle.

This work bridges the fields of mechanics and crystallography, offering a robust framework for understanding mechanical couplings related to symmetry breaking and spatial anisotropy in metamaterial design. It provides a foundational platform analogous to crystal physics and crystal chemistry, paving the way for new innovations in the design of advanced metamaterials

## Acknowledgments

This research is supported by the National Natural Science Foundation of China (Grant no. 12272225), the Ministry of Science and Technology in China (Grant no. SQ2022YFE010363), and the Research Incentive Program of Recruited Non-Chinese Foreign Faculty by Shanghai Jiao Tong University

## Author contributions

D. Zhong and J. Ju designed the research; D. Zhong designed and analyzed the lattices and conducted continuum modeling and programming; D. Zhong performed the FE simulations; D. Qi modeled the prototypes and conducted the experiments; D. Zhong wrote the original draft; J. Ju wrote the paper with review and editing.